RF amplification property of the MgO-based magnetic tunnel junction using field-induced ferromagnetic resonance

K. Konishi[1], D. K. Dixit[2], A. A. Tulapurkar[2], S. Miwa[1], T. Nozaki[3], H. Kubota[3], A. Fukushima[3], S. Yuasa[3] and Y. Suzuki[1]

1 Graduate School of Engineering Science, Osaka University, Toyonaka, Osaka 560-8531, Japan

2 Department of Physics, Indian Institute of Technology, Bombay, Powai, Mumbai 400076, India

3 National Institute of Advanced Industrial Science and Technology (AIST), Spintronics Research Center, Ibaraki 305-8568, Japan

Abstract

 The radio-frequency (RF) voltage amplification property of a tunnel magnetoresistance device driven by an RF external-magnetic-field-induced ferromagnetic resonance was studied. The proposed device consists of a magnetic tunnel junction (MTJ) and an electrically isolated coplanar waveguide. The input RF voltage applied to the waveguide can excite the resonant dynamics in the free layer magnetization, leading to the generation of an output RF voltage under a DC bias current. The dependences of the RF voltage gain on the static external magnetic field strength and angle were systematically investigated. The design principles for the enhancement of the gain factor are also discussed.

The discovery of giant tunneling magneto-resistance (TMR) in MgO-based magnetic tunnel junctions (MTJs) [1,2] accelerates the development of spin devices using MTJs, such as reading heads for hard disk drives (HDDs) and spin-transfer torque magnetic random access memories (STT-MRAMs) [3]. Further, spin-torque oscillators [4] and spin-torque diode effect [5] using MgO-based MTJs have attractive characteristics because of their large magneto-resistance effect. In addition to these devices, a MgO-based MTJ can be used to amplify radio-frequency (RF) signals with frequency tunability. Thus far, several RF amplification properties of an MTJ structure have been proposed and demonstrated experimentally. The study by Slonczewski [6] was the first to propose the above mentioned properties. The other concepts of RF amplification have been proposed using negative differential resistance [7], vortex-core resonance [8,9], and spin-torque-induced ferromagnetic resonance in an MTJ [10]. These devices for RF amplification were combined with the spin-transfer torque effect [11,12]. On the other hand, in this study, we proposed a spin transistor using an MTJ, which is driven by a current-induced magnetic field, and we demonstrated a direct current (DC) power gain of more than 1 at room temperature [13]. In this paper, we present the RF amplification properties in the current-induced magnetic field driven spin transistor.

A schematic structure of the proposed device with the measurement circuit is shown in Fig. 1(a). When a coplanar waveguide (CPW) is combined with an MTJ, an RF magnetic field is generated around the CPW under an application of RF current to the CPW. Further, ferromagnetic resonance (FMR) is induced when the frequency of the RF magnetic field is tuned to the resonant frequency of magnetization of the free layer in the MTJ. Generally, magnetization precession can be excited efficiently under the resonance condition, leading to the generation of a large RF output voltage from the MTJ under a large DC bias current. If the output voltage exceeds the input voltage, the proposed device can be used for the amplification of RFs.

MTJ films with a structure of buffer layers/PtMn/CoFe/Ru/CoFeB(3)/MgO (1.1)/CoFeB(3)/Ru(1.5)/NiFe(2)/capping layers (nm in thickness) were deposited on Si/SiO$_2$ substrates using a magnetron sputtering method (Canon ANELVA C7100). From the current-in-plane tunneling (CIPT) measurements, the resistance-area product and the magneto-resistance ratio were evaluated to be 3.8 $\Omega\mu m^2$ and 110%, respectively. The multilayer film was patterned into junctions (0.3 × 0.6 $\mu m^2$) with an electrically isolated coplanar waveguide. The RF power, which generates the RF magnetic field, was applied at port-1 ($V_{in}$) of the vector network analyzer (VNA). Further, a bias current was applied to the MTJ through a DC port of the bias Tee. The output signal from the MTJ ($V_{out}$) was detected by port-2 of the VNA. The RF amplification property was evaluated by monitoring the $S_{21}$ (= $V_{out}/V_{in}$) parameter. The magnetization configurations of the pinned and free layers and the magnetic field are shown in Fig. 1(b). The $x$-axis (direction perpendicular to the CPW) is defined to be parallel to the easy axis of the free layer. The external magnetic field, $H_{ext}$, is applied along the direction with angle $\theta_{field}$ from the $x$-axis. The $S_{21}$ parameter was measured under various $H_{ext}$ and $\theta_{field}$ conditions. The input RF power and the DC bias current were fixed at 20 $\mu$W and -8 mA in all the measurements. Here, the positive current is defined as one in which the electrons flow from the free layer to the pinned layer.

In order to extract the FMR signal cleanly, the background signal, which originates from the transmission property of the CPW, was subtracted from the raw $S_{21}$ data. The background signal was obtained under the sufficiently large $H_{ext}$ = 1 kOe applied along the direction of the $x$-axis, because the FMR could not be induced in this configuration.

In this paper, $\Delta S_{21}$ indicates the measured $S_{21}$ parameter without the background.

Fig. 2 (a) shows the typical MR curve when $\theta_{\text{field}}$ is 0°. The obtained MR ratio became three times smaller than that obtained by the CIPT measurement, owing to the influence of the parasitic series resistance. A shift field of 50 Oe was obtained, which is caused by orange-peel coupling (a parallel state is preferred under zero magnetic field). The measured gain factor (i.e., the absolute value [$\Delta S_{21}$], which is hereafter simply called $\Delta S_{21}$) as a function of frequency is shown in Fig. 2 (b), where $\theta_{\text{field}}$ was fixed at 55°. Depending on the strength of the external magnetic field, the amplitude of $\Delta S_{21}$ and the resonant frequency were changed, indicating that the FMR excitation was induced by the application of the RF signal to the CPW. The value of $\Delta S_{21}$ becomes small when $H_{\text{ext}}$ = 50 Oe, possibly because of the inhomogeneous precession motion attributed to multi-domain formation. The observed resonant frequency was plotted as a function of the external magnetic field with $\theta_{\text{field}}$ = 55° in Fig. 2 (c). The observed shift was well reproduced by Kittel's formula (red curve in the same figure). From the fitting, the in-plane and out-of-plane anisotropy fields were evaluated to be 15 Oe and 7500 Oe, respectively.

Fig. 3(a) shows $\Delta S_{21\_\text{max}}$ as functions of the strength and angle of $H_{\text{ext}}$. Here, we define $\Delta S_{21\_\text{max}}$ as the peak height of $\Delta S_{21}$ at each angle and external magnetic field. The red area indicates the high amplification gain, which reaches the maximum value of approximately 0.07.

The output voltage from the MTJ ($V_{\text{out}}(\omega)$) for small precession angles is described by eq. (1) [see supplement in ref. 4].

$$V_{out}(\omega) = \eta'(\omega) \frac{R_{\text{AP}} - R_{\text{P}}}{4\sqrt{2}} i_0 (\theta_{\text{prec}}) \sin(\theta_{\text{tilt}}) \quad (1)$$

where $\eta'(\omega)$ is the efficiency of the RF circuit; $R_{\text{P}}$ and $R_{\text{AP}}$ are the resistances in parallel and anti-parallel states, respectively; $i_0$ is the bias current; $\theta_{prec}$ is the precession angle; and $\theta_{tilt}$ denotes the relative angle between the pinned and free layer magnetization. When $\theta_{field} = 55^o$ and $H_{ext} = 100$ Oe, a relative angle of $\theta_{tilt} = 83°$ was obtained by resistance monitoring. The largest RF gain factor was obtained at this field and angle. This behavior is consistent with eq. (1). Furthermore, the resonant frequency decreases with a decrease in the external magnetic field is small, which in turn results in an increase in the precession angle. From the above discussion, it can be concluded that it is important to reduce the resonant frequency and obtain a relative angle of 90°to obtain a large output.

Here, we compare the experimental results with the simulation results. Using the parameters obtained from our experiments and from Kittel's fitting, we simulated the RF gain by using a macro-spin model based on the Landau-Lifshitz-Gilbert equation. Fig. 3(b) shows the simulated result as functions of $H_{ext}$ and $\theta_{field}$. We assume that the damping factor $\alpha = 0.01$ and temperature $T = 300$ K. Here, the influence of the thermal effect is treated as a random magnetic field [15]. The experimental results were well reproduced qualitatively by this simple simulation. However, in the simulation, the maximum gain was obtained for both a small magnetic field and a small tilt angle at around $H_{ext} = 50$ Oe and $\theta_{field} = 15°$. At this field and angle, the shift field and in-plane magnetic anisotropy are cancelled simultaneously by the external magnetic field. Subsequently, the resonant frequency becomes very low and the RF gain increases. However, this condition is difficult to achieve in the experiment, because a multi-domain structure is formed in the low magnetic field range. Further, the simulated gain is approximately 5 times greater than that obtained by the experimental result. In the experiment, the linewidth of the FMR peak is broadened because of non-uniform precession attributed to a relatively large sample size. This leads to a small Q-value (~2) in the experiment, and consequently, the gain factor becomes smaller than expected.

The RF gain does not exceed 1 even in the simulation. One possible approach to realize high RF amplification is to introduce perpendicular magnetic anisotropy [16]. When the out-of-plane anisotropy is cancelled by the external magnetic field, a high Q-value of 50 can be obtained. This value is 25 times greater than that obtained by our results. Thus, the RF gain factor can exceed 1 if we apply their condition (i.e., using Fe-rich CoFeB free layer for perpendicular magnetic anisotropy and applying external magnetic field to cancel demagnetization field) to our experiment.

Another possible approach to enhance the RF gain factor is to utilize the spin transfer effect. In the present measurement conditions, the spin transfer torque has only a small influence on the RF gain, because $\theta_{tilt}$ is nearly 90 degree when the maximum $\Delta S_{21}$ was obtained and the electric breakdown voltage of the sample was relatively low. However, if the effect of the spin torque is sufficiently large, the effective damping factor will be reduced [14], inducing a large precession angle; hence, the output voltage from the MTJ can be increased, leading to the realization of a higher RF gain. This technique can be adapted to a magnetic field feedback oscillator [15].

In summary, the amplification property of MTJs afforded by the magnetic-field-induced FMR was proposed and demonstrated. A maximum voltage gain of 0.07 was achieved under the optimized external magnetic field condition. In addition, the static external magnetic field strength and angle dependences of the voltage gain were systematically investigated and reproduced qualitatively by a simple macro-spin model simulation. The improvement of the uniform precession in the element, as well as the introduction of perpendicular magnetic anisotropy or the spin transfer effect, were shown to be effective in achieving RF voltage gains greater than 1.

Acknowledgement

This work was mainly supported by the New Energy and Industrial Technology Development Organization (NEDO) Spintronics Nonvolatile Device Project. K. Konishi is supported by a Research Fellowship of the JSPS for young scientists.

Figure 1

(a)

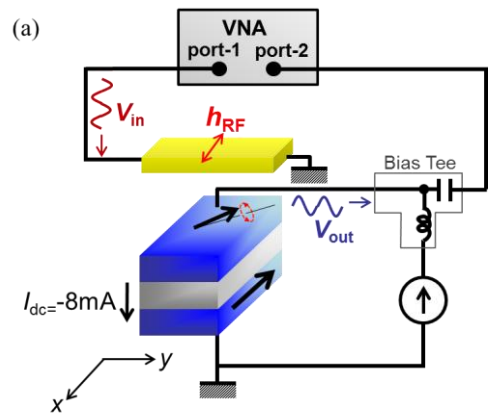

(b)

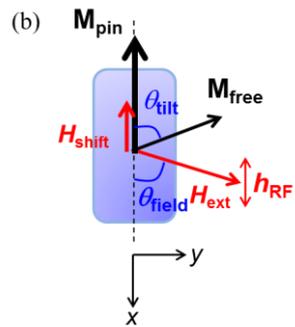

(c)

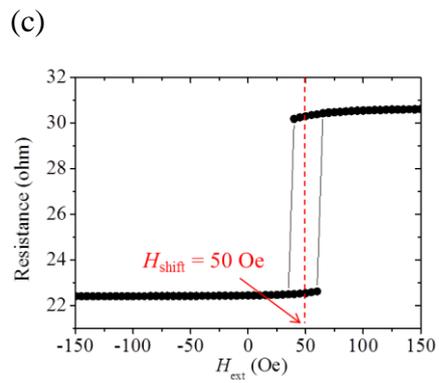

Figure 2

(a)

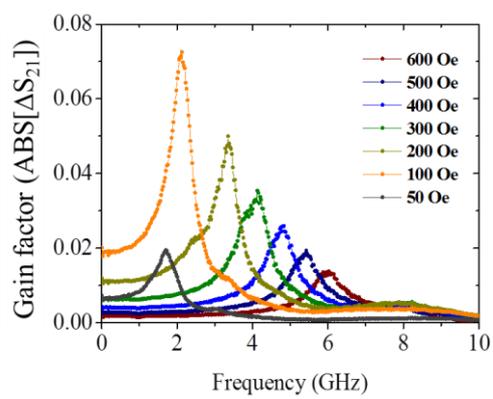 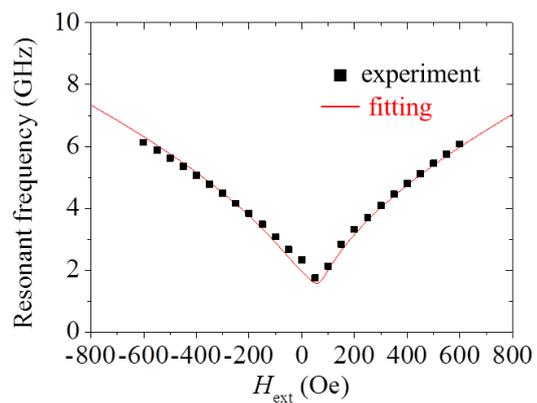

(b)

Figure 3

(a)
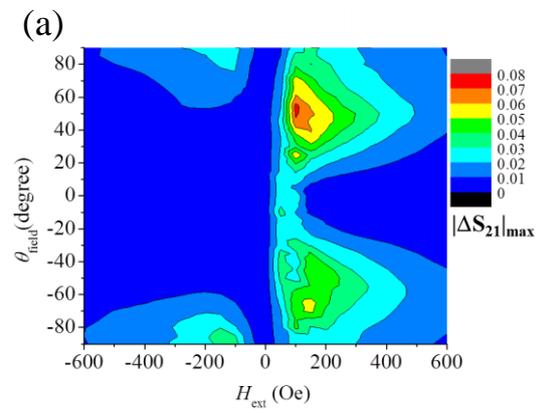

(b)
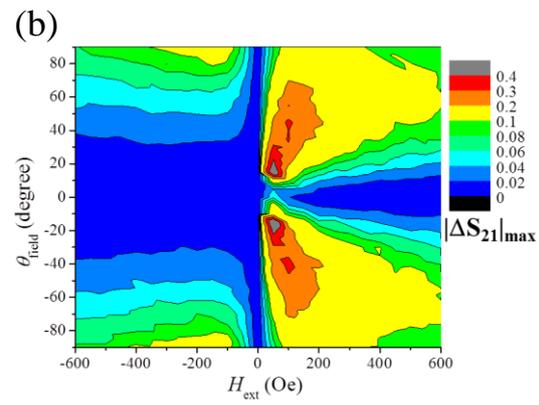

Figure captions

Figure 1

(a) Concept of the proposed device with the measurement circuit. The FMR in the free layer of the MTJ is induced by an RF magnetic field due to the application of an RF signal from port-1 of the VNA. Under DC bias, the MTJ produces an RF output, which is applied to port-2 of the VNA. The VNA measures the $S_{21}$ (= $V_{out}/V_{in}$) parameter, which represents the gain factor.

(b) Schematic image of the configuration of magnetization and magnetic fields. $M_{free}$ and $M_{pin}$ denote the magnetization of the free and pinned layer; $\theta_{tilt}$ and $\theta_{field}$ indicate the relative angle and the angle of the external magnetic field; and $h_{RF}$, $H_{ext}$, and $H_{shift}$ indicate the RF magnetic field, the external magnetic field, and shifted field, respectively.

Figure 2
(a) Typical MR curve ($\theta_{field} = 0°$).

(b) Gain factor (absolute value [$\Delta S_{21}$]) as a function of frequency ($\theta_{field} = 55°$).

(c) Resonant frequency as a function of the external magnetic field ($\theta_{field} = 55°$). The black dots and red curve show the experimental results and the fitting curve, respectively.

Figure 3
$S_{21\_max}$ as functions of $\theta_{field}$ and $H_{ext.}$ (a) Experimental result and (b) simulated result.